\documentstyle[preprint,eqsecnum,epsfig,prd,aps,floats]{revtex} 

\begin{document}

\newcommand{\drawsquare}[2]{\hbox{%
\rule{#2pt}{#1pt}\hskip-#2pt
\rule{#1pt}{#2pt}\hskip-#1pt
\rule[#1pt]{#1pt}{#2pt}}\rule[#1pt]{#2pt}{#2pt}\hskip-#2pt
\rule{#2pt}{#1pt}}

\newcommand{\Yfund}{\raisebox{-.5pt}{\drawsquare{6.5}{0.4}}}
\newcommand{\Ysymm}{\raisebox{-.5pt}{\drawsquare{6.5}{0.4}}\hskip-0.4pt%
        \raisebox{-.5pt}{\drawsquare{6.5}{0.4}}}
\newcommand{\Yasymm}{\raisebox{-3.5pt}{\drawsquare{6.5}{0.4}}\hskip-6.9pt%
        \raisebox{3pt}{\drawsquare{6.5}{0.4}}}
\newcommand{\Ythree}{\raisebox{-3.5pt}{\drawsquare{6.5}{0.4}}\hskip-6.9pt%
        \raisebox{3pt}{\drawsquare{6.5}{0.4}}\hskip-6.9pt
        \raisebox{9.5pt}{\drawsquare{6.5}{0.4}}}

\newcommand{\nl}{\nonumber \\}
\newcommand{\eq}[1]{Eq.~(\ref{#1})}

\newcommand{\Tr}{{\rm Tr}}
\newcommand{\diag}{{\rm diag}}

\preprint{\vbox{ \tighten {
		\hbox{MIT-CTP-2847}
		\hbox{PUPT-1844}
		\hbox{IASSNS-HEP-99/35}
                \hbox{hep-th/9904050}
		}  }}  
  
\title{${\cal N} = 1$ theories, T-duality, and AdS/CFT correspondence} 
  
\author{ Martin Gremm\thanks{email: gremm@feynman.princeton.edu }\thanks{ 
On leave of absence from MIT, Cambridge, MA 02139}} 
  
\address{Joseph Henry Laboratories, Princeton University, Princeton, NJ 08544} 
 
\author{Anton Kapustin\thanks{email: kapustin@ias.edu}} 
 
\address{ School of Natural Sciences, Institute for Advanced Study\\ 
Olden Lane, Princeton, NJ 08540} 
 
\maketitle  
 
\begin{abstract} 
\tighten{ 
We construct an ${\cal N}=1$ superconformal field theory using branes of 
type IIA string theory. The IIA construction is related via T-duality to a  
IIB configuration with 3-branes in a background generated by two intersecting  
O7-planes and 7-branes. The IIB background can be viewed as a local piece of  
an F-theory compactification previously studied by Sen in connection with the 
Gimon-Polchinski orientifold. We discuss the deformations of the IIA and IIB  
constructions and describe a new supersymmetric configuration with curving 
D6-branes. Starting from the IIB description we find the supergravity dual of  
the large $N$ field theory and discuss the matching of operators  
and KK states. The matching of non-chiral primaries exhibits some interesting
new features. We also discuss a relevant deformation of the field theory 
under which it flows to a line of strongly coupled ${\cal N}=1$ fixed points 
in the infrared. For these fixed points we find a partial supergravity description. 
} 
\end{abstract} 
 
\newpage

\section{Introduction} 
 
Brane constructions in string theory provide powerful tools for analyzing
field theories in diverse dimensions and with varying 
amounts of supersymmetry~\cite{branes1,branes2}. For a review and 
references see \cite{review}. More recently the Maldacena conjecture
\cite{m1,gubs,w1} added a new relation between the large $N$  
limit of conformal field theories on branes and the near horizon geometry of 
the corresponding black brane solutions. The original conjecture was stated 
for ${\cal N}=4$ SYM realized on $N$ 3-branes, but subsequently more general 
examples were discovered. One class of such examples are orbifolds of the  
${\cal N} = 4$ configuration \cite{ks,oz} and another includes theories on 
3-branes in nontrivial F-theory backgrounds \cite{f1,f2,kap}. All of these 
constructions give rise to conformal theories with varying amounts of 
supersymmetry. A third class of theories arises on 3-branes at a conifold 
singularity \cite{kw,ur,nek,unge,josh}. These ${\cal N}=1$ theories are not 
conformal at all scales, but flow to a line of conformal fixed points in the 
infrared. For all these theories the correspondence between the large $N$ 
field theory and supergravity was studied in some detail. 
For branes on a conifold it turned out to be useful to have a type IIA description which is 
related to the IIB configuration via T-duality~\cite{ur}. 
 
In this paper we study a superconformal ${\cal N} = 1$ $Sp(N)\times Sp(N)$ 
gauge theory with matter in the fundamental, bifundamental, and antisymmetric 
representations. We also discuss a specific 
deformation which preserves ${\cal N}=1$ SUSY but breaks conformal invariance.  
The resulting theory has a running gauge coupling and flows to a line of  
superconformal fixed points in the infrared. 
For both of these theories we give a IIA brane construction as well as a IIB  
orientifold construction. The latter description allows us to obtain the supergravity  
solution that is dual to the large $N$ limit of the conformal field theory. 
The type IIA description, on the other hand, provides a simple way to determine 
the gauge group, the matter content, and the superpotential of the theories in 
question.  
 
In most ${\cal N}=1$ theories discussed in the AdS/CFT literature (see e.g.
\cite{ks,kw}) 
the R-current which is the superpartner of the stress-energy tensor can be
fixed uniquely 
by field theory considerations. For the theories we discuss here this is not
the case. There 
is a one parameter family of candidate R-currents, both in the theory with 
vanishing beta function, and in its deformation which flows to a line of fixed
points. 
Since the R-charges of the fields are not uniquely determined, there is no
field theory prediction for the 
dimensions of the chiral primary operators. On the other hand, once we have a  
supergravity dual of the large $N$ field theory, we know which gauge boson on
AdS is the superpartner of the graviton. If we are able to match 
field theory operators
with supergravity states, we can determine the R-charges 
of all fields and therefore the dimensions of all chiral primary operators.

Although there is no firm field-theoretical prediction for the dimensions of fields
in the  infrared, for the theory with vanishing beta function the most natural
assumption is that all fields have canonical dimensions, i.e., that the theory
is finite. This will  
be born out by the supergravity analysis. In the other case, the theory with 
a running coupling constant, the correct charge assignment in the 
infrared is harder to guess. Unfortunately the supergravity analysis in this case is  
on a considerably less solid footing and depends on circumstantial evidence. 
Nonetheless our analysis suggests a definite R-charge assignment. It would be interesting 
to find a field theory explanation for it.

The type IIA construction involves D4-branes compactified on a circle 
as well as NS5-branes, D6-branes, and O6-planes. The gauge theory lives 
on the D4-branes. Our construction is very similar to the  
brane configurations that give rise to elliptic ${\cal N}=2$ models 
\cite{branes2,angfinite,jaemo,kap}. One advantage of the IIA description is that the 
moduli space of the gauge theory is realized geometrically. The flat directions 
correspond to motions of the 4-branes. Similarly, relevant perturbations 
of the field theory, such as masses for the matter fields, are also realized 
geometrically as motions of the 6-branes. This allows us 
to identify a 6-brane configuration that gives rise to a superconformal 
${\cal N}=1$ theory on the 4-branes with an exactly marginal parameter.  
We can also identify relevant perturbations of the  
superconformal 4-brane theory that lead to theories with running coupling 
constants. There is one particular perturbation that gives rise to a theory 
that flows to a line of conformal fixed points in the infrared. 
The moduli space of the perturbed theory has a Coulomb branch. A generic 
${\cal N} = 1$ theory with a Coulomb branch has a low 
energy effective gauge coupling that varies over the moduli space. The theory 
we are considering in this paper has the special feature that the low-energy effective 
gauge coupling does not depend on the moduli. This will be relevant 
when we discuss the supergravity description of these theories. 
 
In order to construct the supergravity duals we  
T-dualize the IIA configuration along the compact direction. This operation 
turns the D6-branes and the O6-planes into D7-branes and O7-planes. The D4-branes 
turn into D3-branes probing this background. Similar probe theories were
studied 
in \cite{bds,jhs,aks,ah1}, and their relation to supergravity is described in 
\cite{f1,f2,kap}. Our IIA configuration turns out to be T-dual to 3-branes 
probing a local 
piece of an F-theory compactification \cite{sen1,sen2} which is related to 
the Gimon-Polchinski model \cite{gp}. The simplest such configuration, 
consisting of two intersecting O7-planes with four coincident 7-branes
on top of each, corresponds to the IIA construction of the superconformal  
4-brane theory. In the  type IIB construction the Ramond-Ramond (RR) charges of the 
7-branes are cancelled locally by the charges of the orientifold planes, so the string 
coupling is constant. Since the type IIB description is a perturbative orientifold, 
we can find the supergravity dual of the large $N$ limit of the field theory 
along the lines of \cite{f1,f2}. Matching the spectrum of primary operators
with the KK modes allows us to determine the $U(1)_R$ charges of 
all fields in the conformal theory unambiguously. The matching of non-chiral
primaries exhibits a new interesting feature: we find a short supergravity
multiplet whose field theory counterpart becomes short only when $N\to\infty$.
We interpret this as the evidence that at higher orders in $1/N$ supersymmetry
mixes one-particle and two-particle supergravity states. 
 
It should also be possible to find a supergravity description of the
infrared limit of the deformed theory. Although this theory is not conformal, it
has a constant low-energy  
effective coupling along the Coulomb branch, so the supergravity dual will have a
constant dilaton. To find this dual we need to study the  
deformations of the backgrounds in IIA and IIB and find an explicit map between them.  
As mentioned before, this is straightforward on the IIA side, since the deformations  
correspond to motions of the 6-branes. On the IIB side the situation is more 
involved. We can analyze the deformations on the IIB side by studying the
theory on the 7-branes. 
The eight-dimensional theory on the 7-branes has six-dimensional matter  
localized at the intersection of orthogonal 7-branes. We analyze the moduli 
space of this impurity theory following \cite{kap2}, and find an 
explicit map between the type IIB and type IIA deformations.  
Among supersymmetric type IIB deformations there is one that maps to a new IIA 
brane configuration which involves curving D6-branes in the background of an 
NS5-brane. The map between deformations also allows us to identify the IIB
configuration that gives rise to the non-conformal probe theory 
with moduli-independent effective coupling. We do not have a complete supergravity 
description of this theory, but a partial description is possible. It supplies enough 
information to determine the dimension of all chiral operators in the infrared if 
we use field theory considerations as well.

In section \ref{IIA} we discuss the type IIA construction of the probe theory 
and list some field theory results that we need in subsequent sections. 
Section \ref{IIB} contains the T-duality, the analysis of the 7-brane 
impurity theory, and the map between IIA and IIB deformations. We also 
briefly discuss the exotic IIA deformation that appears as the counterpart of 
an ordinary deformation in IIB. In section \ref{sugra} we analyze the large $N$ 
limit of our field theories and their supergravity duals. We discuss the matching 
of operators with Kaluza-Klein modes in the conformal case and present a 
partial analysis in the non-conformal case.

\section{The IIA construction of the field theory} 
\label{IIA} 
 
\subsection{The IIA brane configuration} 
 
\begin{figure} 
\centerline{\epsfxsize = 11truecm \epsfbox{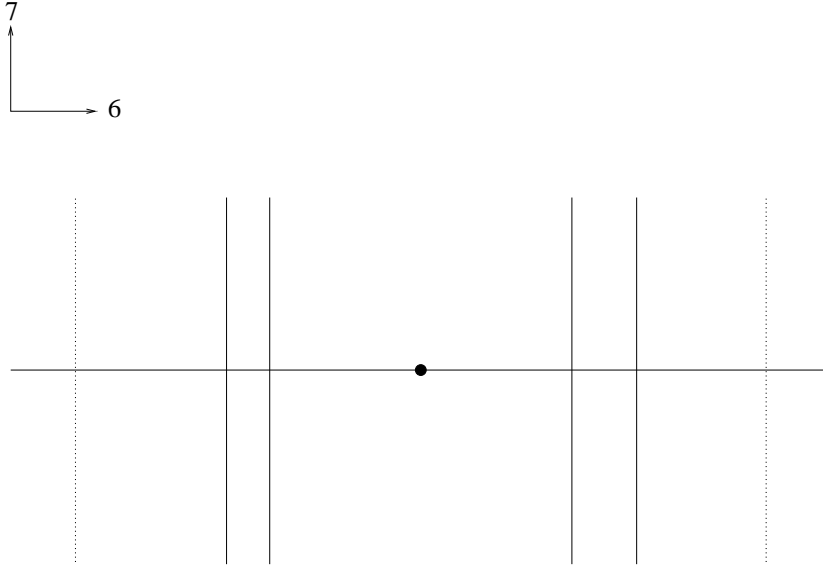} } 
\tighten{ 
\caption{ Brane configuration: The vertical dashed lines are the 
O6-planes, the solid lines are the D6-branes, the horizontal line are the 
D4-branes, and the point represents the NS5-brane. Only half of the $X_6$ 
circle is shown.} 
\label{fig1} 
} 
\end{figure}

A configuration consisting of D4-branes extending in 01236, D6-branes and
O6-planes extending
in 0123789, and NS5-branes extending in 012389 preserves four supercharges.
We obtain an 
${\cal N}=1$ supersymmetric field theory in four dimensions after compactifying 
$X_6$ on a circle with circumference $2\pi R_6$. Specifically we consider
configurations 
with $N$ D4-branes wrapping the compact $X_6$ direction. We put two
O6$^-$-planes at 
$X_6=0,\pi R_6$ and an NS5-brane and its image at $X_6 = R_6\pi/2,3R_6\pi/2$. In
order to cancel 
the total RR charge, we place four physical D6-branes on the circle.  
An example of such a configuration is shown in Fig.~\ref{fig1}. 
 
These brane configurations are very similar to the configurations that give
rise to finite 
${\cal N} =2$ theories in four dimensions \cite{angfinite,jaemo,kap}.  
In fact, the configuration we study here can be obtained from one of the ${\cal N} =2$ 
configurations in \cite{angfinite} by rotating the NS5-branes from the 45 directions into 
the 89 directions. This breaks half of the supersymmetries, giving an ${\cal N}=1$ 
theory in $d=4$. 
 
Using standard techniques \cite{review}, we can determine the matter content 
and the superpotential of the field theory on the 4-branes.  
Unlike the ${\cal N} =2$ case, the $X_6$ position of the D6-branes will 
play an important role in our analysis. We need to distinguish two cases 
that are of interest for the analysis in this paper. 
Either all 6-branes intersect the NS5-brane, 
or the four 6-branes are split into two groups of two to the left and right 
of the NS5-brane (as shown in Fig.~\ref{fig1}). These two  choices give rise 
to physically inequivalent theories. The former configuration yields a line 
of fixed points (parametrized by the dilaton expectation value) that passes
through zero coupling, 
while the latter corresponds to a non-conformal gauge theory which flows to 
a line of strongly coupled fixed points. 
 
\subsection{The conformal case: a field theory analysis}\label{conformal}
\label{iib}

The theory on the 4-branes turns out to be an $Sp(2N)_1\times Sp(2N)_2$ gauge 
theory with matter fields $A_i, i=1,2$ in the antisymmetric representation  
of each of the gauge groups, two bifundamentals ${\cal Q}, \tilde{\cal Q}$, 
and fundamentals from the 4-6 strings. The brane configuration, and consequently 
the field theory, admit a symmetry which exchanges the two $Sp$ factors. To
determine the number and flavor 
representations of the fundamentals we need to understand the classical  
gauge theory on the 6-branes. Note that the worldvolume of the NS5-brane  
lies within the worldvolume of the D6-branes. It was argued in \cite{karch} 
that the 6-branes can break on the NS5-brane (see also \cite{nscft}). The gauge
group on the four 6-branes turns out to be $U(4)_u\times U(4)_d$, where the
two $U(4)$ factors act on the upper and lower halfs of the 6-branes respectively.
One-loop effects break the $U(4)_u\times U(4)_d$ symmetry to
$SU(4)_u\times SU(4)_d$~\cite{berkoozetal}.
The matter content of the 6-brane theory includes a bifundamental 
hypermultiplet from strings connecting upper and lower halfs of the 6-branes. 
We will have more to say about the 6-brane theory when we discuss the 
deformations of this background. For our present purposes we only need to
know that the gauge group of 
the 6-brane theory is the flavor group of the probe theory. 
 
The matter content and the superpotential  
for a 4-brane probe in this background were worked out in \cite{karch}. 
The fundamentals transform as $q=(\Yfund,1,{\bf 4},1)$, 
$\tilde{q}=(\Yfund,1,1,{\bf 4})$, 
$p=(1,\Yfund,{\bar {\bf 4}},1)$, and $\tilde{p}=(1,\Yfund,1,{\bar {\bf 4}})$ 
under 
$Sp(2N)_1\times Sp(2N)_2\times SU(4)_u \times SU(4)_d$. The superpotential 
reads 
\begin{equation} 
W = h_1\tilde{\cal Q}A_1 J_1{\cal Q} - h_1{\cal Q}A_2 J_2\tilde{\cal Q} + 
h_2q{\cal Q} p + h_2\tilde{p}\tilde{\cal Q}\tilde{q}. 
\end{equation} 
Here $J_1$ ($J_2$) is the invariant antisymmetric tensor of $Sp(2N)_1$ ($Sp(2N)_2$). 
Following \cite{line} it is easy to check that this theory has a line of  
fixed points passing through weak coupling. The one-loop beta function vanishes
and the symmetry between the gauge factors implies that both antisymmetric 
tensors have the same anomalous dimension $\gamma_A$, both bifundamentals 
have $\gamma_{\cal Q}$ and all fundamentals have $\gamma_q$. Therefore   
the beta functions of the gauge coupling and the 
Yukawa couplings in the superpotential are
\begin{eqnarray}\label{beta1}
\beta_g &\sim& 2(N-1)\gamma_A+4N\gamma_{\cal Q}+8\gamma_q \nl 
\beta_{h_1}&\sim& 2\gamma_{\cal Q}+\gamma_A\\ 
\beta_{h_2}&\sim& \gamma_{\cal Q}+2\gamma_q. \nonumber 
\end{eqnarray}
Setting all beta functions to zero gives two independent constraints on the  
three coupling constants. The remaining coupling constant parametrizes a  
line of superconformal fixed points. Since setting all anomalous dimensions 
to zero satisfies the constraints, this line passes through the free point  
$g=h_1=h_2=0$. Note that requiring the beta functions to vanish does not 
fix anomalous dimensions unambiguously. The most natural assumption is
that the dimensions of the fields are unchanged as one moves along the fixed 
line. This would mean that the theory is finite. 
The supergravity computation in the last section supports this
conjecture by showing that this is true in the large $N$ limit.
 
The moduli space of this theory includes subspaces where it flows to 
theories with more supersymmetry. For example, giving an expectation value 
to either ${\cal Q}$ or $\tilde{\cal Q}$ proportional to a unit matrix gives a  
mass to half of the fundamentals and breaks the gauge group to the diagonal  
$Sp(2N)_D$. It is a simple matter to show that the resulting theory flows to an 
${\cal N}=2$ superconformal theory with gauge group $Sp(2N)$, one antisymmetric  
hypermultiplet, and four hypermultiplets in the fundamental. Giving such
expectation values to both ${\cal Q}$ and $\tilde{\cal Q}$ makes all flavors massive  
and breaks the gauge group to $SU(N)$. Part of the bifundamentals are eaten by gauge  
bosons, and the rest give rise to three chiral superfields in the adjoint of $SU(N)$.  
This theory flows to ${\cal N} = 4$ SYM in the infrared. 
 
These field theory results are reproduced in the brane construction  
if we identify the positions of the D4-branes with the field theory moduli 
in the following way: 
\begin{eqnarray} 
&X_7 \sim {\cal Q Q}^\dagger - \tilde{\cal Q}^\dagger \tilde{\cal Q} & \\ 
&X_4+iX_5 \sim {\cal Q}\tilde{\cal Q}.& 
\end{eqnarray} 
Giving an expectation value to either of the bifundamentals while keeping the other  
expectation value zero 
corresponds to moving the 4-branes in the positive or negative $X_7$ direction. 
Turning on both bifundamentals corresponds to moving the 4-branes in the $X_4$ 
and $X_5$ directions as well as $X_7$. The effect of these motions on the 4-brane theory 
agrees with the field theory expectations. If we move the 4-branes off the NS5-branes in  
the $X_7$ direction, we can ignore the NS5-brane. The remaining branes preserve eight  
supercharges, and standard techniques \cite{review} confirm the matter content and gauge 
group stated above for the ${\cal N}=2$ case. Moving the 4-branes in $X_4$ 
and $X_5$ amounts to separating them from all other branes. The theory on 
the 4-branes is then ${\cal N} = 4$ SYM as expected. 
 
\subsection{The non-conformal case}\label{iic}
We can deform the background for the 4-brane theory by moving the 6-branes 
in the $X_6$ and $X_{4,5}$ directions. These brane motions are parametrized 
by expectation values of the two complex scalars, ${\cal M}, \tilde{\cal M}$ 
in the bifundamental hypermultiplet of the $(1,0)$ theory on the intersection of 
the 6-branes and the NS5-branes \cite{karch}. 
More precisely, we relate the positions of the 6-branes to
${\cal M}, \tilde{\cal M}$ as follows:
\begin{eqnarray} 
&X_6 \sim {\cal M M}^\dagger - \tilde{\cal M}^\dagger \tilde{\cal M} & \\ 
&X_4+iX_5 \sim {\cal M}\tilde{\cal M}.& 
\end{eqnarray} 
To obtain the configuration shown in Fig.~\ref{fig1} we have to set 
${\cal M}=\diag(m_1,m_2,0,0)$ and 
$\tilde{\cal M}=\diag(0,0,\tilde{m}_3,\tilde{m}_4)$. 
These bifundamental expectation values act as mass terms in the 4-brane theory.
The corresponding terms in the field theory superpotential are 
\begin{equation} 
W = \tilde{\cal M}q\tilde{q} + {\cal M}p\tilde{p}. 
\end{equation} 
We will be particularly interested in the case
$m_1=m_2=\tilde{m}_3=\tilde{m_4}$.  
In this case the bifundamental expectation values break the
$SU(4)_u\times SU(4)_d$ 6-brane gauge group  
to $SU(2)_1\times SU(2)_2\times U(1)$. 
After integrating out the massive components of the fundamentals, the  
superpotential of the 4-brane theory reads 
\begin{equation}\label{defW} 
W = h_1\tilde{\cal Q}A_1 J_1{\cal Q} - h_1{\cal Q}A_2 J_2\tilde{\cal Q} + 
h_3 q{\cal Q}\tilde{\cal Q}\tilde{q} + h_3 \tilde{p}\tilde{\cal Q}{\cal Q}p. 
\end{equation} 
The fundamentals now transform as $q=(\Yfund,{\bf 1},{\bf 2},{\bf 1})$, 
$\tilde{q}=(\Yfund,{\bf 1},{\bf 2},{\bf 1})$, $p=({\bf 1},\Yfund,{\bf 1},{\bf 2})$, and 
$\tilde{p}=({\bf 1},\Yfund,{\bf 1},{\bf 2})$ under 
$Sp(2N)_1\times Sp(2N)_2\times SU(2)_1 \times SU(2)_2$. Actually, the  
superpotential, \eq{defW}, has an accidental $SO(4)_1\times SO(4)_2$ global symmetry 
under which $q$ and $\tilde{q}$ transform as a $({\bf 4}, {\bf 1})$ while 
$p$ and $\tilde{p}$ transform as $( {\bf 1},{\bf 4})$.

An analysis along the lines of \cite{line} shows that this theory also has a 
line of superconformal fixed points. The beta functions are given by 
\begin{eqnarray}\label{beta2} 
\beta_g &\sim& 4+2(N-1)\gamma_A+4N\gamma_{\cal Q}+4\gamma_q \nl 
\beta_{h_1} &\sim& 2\gamma_{\cal Q}+\gamma_A\\ 
\beta_{h_3} &\sim& 1+\frac{1}{2}\gamma_{\cal Q}+\gamma_q. \nonumber 
\end{eqnarray} 
Demanding that the beta functions vanish, we again find that two out of the 
three constraints are independent, leaving us with a line of fixed points. 
In this case, however, the line does not pass through weak coupling, since at  
least one of the anomalous dimensions must be nonzero. Again the vanishing of
the beta functions alone does not determine the values of anomalous dimensions.
In the  last section we will argue that supergravity considerations allow us  
to fix this ambiguity for large $N$ and find $\gamma_A=\gamma_Q=0,\gamma_q=-1$.
 
As in the conformal case we can analyze the RG flows both in field 
theory and using the brane picture. From the brane construction it is clear 
that we flow to the same ${\cal N}=2$ theory as in the conformal case if we move  
the 4-branes off the NS5-brane in the positive or negative $X_7$ direction. Moving the  
4-branes in $X_{4,5}$ again yields ${\cal N}=4$ SYM. The analysis in the field 
theory is a little more involved in this case because the one-loop 
beta function does not vanish. This implies that there will be threshold 
effects in the matching of the running gauge coupling. On general grounds
one would expect the low-energy effective coupling to depend on the size of the 
bifundamental expectation values in the field theory. However, if we give arbitrary 
(non zero) expectation values to ${\cal Q}$ and $\tilde{\cal Q}$, fields get integrated 
out at a variety of scales. Assuming that the expectation value of ${\cal Q}$ is larger  
than that of $\tilde{\cal Q}$, the $Sp(2N)\times Sp(2N)$ gauge group is broken to the 
diagonal group at a scale set by ${\cal Q}$. The diagonal $Sp(2N)_D$ is 
broken to $SU(N)$ at a scale set by $\tilde{\cal Q}$, and finally the  
fundamentals are integrated out at scale $h_3 {\cal Q}\tilde{\cal Q}$. Matching 
the gauge couplings at each of these scales we find that the low-energy  
effective coupling does not depend on the bifundamental expectation values. This is a  
special feature of this theory that will be important later on.  
 
\section{The type IIB description} 
\label{IIB} 
\subsection{T-duality}\label{iia} 
 
In this section we describe the IIB configuration which is obtained by T-dualizing
the IIA brane configuration of section II along $X_6$. Since $\partial /\partial X_6$
is not a Killing vector, performing this T-duality is not completely trivial.
Similar T-dualities on IIA configurations that  
preserve ${\cal N}=2$ supersymmetry on the 4-branes have appeared in the 
literature \cite{kap,jaemo}. In the ${\cal N}=2$ case the T-duality maps 
the two O6$^-$-planes and the four D6-branes to an orientifold 7-plane and four 
D7-branes. The D4-branes become D3-branes probing this background. The NS5-brane 
and its mirror image turn into a ${\bf Z}_2$ orbifold acting on the 7-brane coordinates  
transverse to the D3-brane. The T-dual of the IIA configuration without NS5-branes was  
analyzed in \cite{bds,ah1}. 
 
Our configuration differs from the ${\cal N}=2$ case by the orientation of the 
NS5-branes. Since this modifies the T-duality considerably we discuss it in 
some detail here. 
 
Our first goal is to T-dualize the NS5-branes and the pair of O6$^-$-planes.  
The other branes can be added later.  
We begin by separating the NS5-brane and its image in the $X_{4,5}$ directions.
The T-dual of the two NS5-branes is a two-center Taub-NUT space.  
Recall that the two-center Taub-NUT space can be thought of as a circle fibered 
over ${\bf R}^3$ so that its radius vanishes at two points on ${\bf R}^3$ (the 
centers). In the present case ${\bf R}^3$ is parametrized by $X_4,X_5,X_7$, while 
the coordinate along the circle is T-dual to $X_6$. The positions of the centers  
correspond to the positions of the NS5-branes in $X_4,X_5,X_7$. 
In the IIA configuration the orientifold projection ensures that position of the physical  
NS5-brane and its image are related by a reflection of the $X_{4,5}$ coordinates. 
The T-dual orientifold projection should therefore impose a similar  
constraint on the location of the centers of the Taub-NUT.  
The Taub-NUT metric has the following form 
\begin{eqnarray} 
ds^2 &=& \left(\frac{4}{b^2}+\frac{1}{R_+}+\frac{1}{R_-}\right)^{-1} 
\left[ d\sigma+ 
\left(\frac{Z_+}{R_+}+\frac{Z_-}{R_-}\right)d\arctan\left(\frac{Y}{X}\right) 
\right]^2 \nl 
&&+\left(\frac{4}{b^2}+\frac{1}{R_+}+\frac{1}{R_-}\right) 
\left[ dX^2+dY^2+dZ^2\right], 
\end{eqnarray} 
where 
\begin{equation} 
Z_\pm = Z \pm Z_0 \qquad R_\pm^2 = X^2+Y^2+Z^2_\pm. 
\end{equation} 
The $\bf{R}^3$ base is parametrized by $X,Y,Z$, the two centers are located at  
$(0,0,\pm Z_0)$, and $\sigma$ is the $4\pi$-periodic coordinate on the circle fiber. 
The parameter $b$ is the asymptotic radius of the fiber. The reflection of 
$X_{4,5}$ in the IIA picture map into reflections of $Z$ and one other  
coordinate of $\bf{R}^3$, say $Y$.  
 
We will be interested in the limit when the asymptotic radius of the circle
fiber, $b$, 
becomes infinitely large, while the T-dual circle parametrized by $X_6$ shrinks
to zero. In this limit the two-center Taub-NUT space becomes an $A_1$ ALE
space, also known as Eguchi-Hanson space. It is useful to change coordinates
\cite{prasad} to transform the metric above into the Eguchi-Hanson form: 
\begin{eqnarray} 
X &=& \frac{1}{8} \sqrt{r^4-a^4} \sin(\theta)\cos(\psi) \\ 
Y &=& \frac{1}{8} \sqrt{r^4-a^4} \sin(\theta)\sin(\psi) \\ 
Z &=& \frac{1}{8} r^2 \cos(\theta)\\ 
\sigma &=& 2\phi, 
\end{eqnarray} 
where $a^2=8Z_0$ and $\psi$ has period $2\pi$.  
The orientifold-induced projection $(Y,Z)\sim (-Y,-Z)$, implies the
identification 
$(\theta,\psi)\sim (\pi-\theta,-\psi)$ for the angular coordinates. The fixed 
locus of this identification is a two-dimensional submanifold of the Eguchi-Hanson 
space which has the topology of a cylinder. Next we want to bring 
the NS5-brane and its image back to the origin of 
the $X_{4,5}$ plane in the IIA description, which corresponds to setting $a=0$. 
For $a=0$ the Eguchi-Hanson metric becomes an orbifold metric on ${\bf C}^2/{\bf Z}_2$. 
To make this explicit we can introduce two complex coordinates 
\begin{equation}\label{z12} 
z_{1,2} = 
r\exp(i\phi/2)\left(\cos(\theta/2)\exp(i\psi/2)\pm 
 i\sin(\theta/2)\exp(-i\psi/2)\right). 
\end{equation}  
In these coordinates the $a=0$ Eguchi-Hanson metric becomes flat. 
The identification $\psi\to \psi+2\pi$ requires that we identify $(z_1,z_2) 
\to (-z_1,-z_2)$ as expected for ${\bf C}^2/{\bf Z}_2$. The additional 
orientifold identification acts on the new coordinates as 
$(z_1,z_2)\to(z_1,-z_2)$, and acting with both orientifold and orbifold
identifications flips the sign of $z_1$. The orientifold projections have two
fixed planes, $z_{1,2}=0$, which we identify with 
two O7$^-$-planes. To summarize, the NS5-brane together with two O6$^-$-planes 
become, under T-duality, a pair of intersecting O7$^-$-planes with six common directions.
 
Now let us put in D-branes. The four physical D6-branes in IIA are located at 
$X_4=X_5=0$. Under T-duality they become D7-branes wrapping the circle fiber of
the Taub-NUT and located at $Y=Z=0$. In other words, they are wrapped on the
invariant cylinder of the orientifold projection. Taking the limit
$b\to \infty, a\to 0$ we 
find that the invariant cylinder develops a neck and becomes a pair of planes 
$z_1=0$ and $z_2=0$ in ${\bf C}^2/{\bf Z}_2$. Thus the four physical D7-branes must be 
located on these planes. Recall that these planes are the O7$^-$-planes and
therefore have 7-brane charge $-4$. It follows that the 7-brane charge is
cancelled between the D7-branes and the orientifold planes, 
and the IIB dilaton is constant. Finally, T-duality turns the D4-branes into
D3-branes extending in 0123. To summarize, the T-dual of the IIA
configuration in the limit  
when the radius of $X_6$ goes to zero consists of  an O7$^-$-plane
with four coincident D7-branes in 01236789, another O7$^-$-plane with 
four coincident D7-branes in 01234589 and 3-branes in 0123. We will refer to 
the 7-branes extending in 01234589 as 7$'$-branes. The orientifold group for  
this configuration is 
\begin{equation}\label{ogroup} 
G = \{ 1, (-1)^{F_L}R_{45}\Omega,(-1)^{F_L} R_{67}\Omega,R_{4567} \}, 
\end{equation} 
where $R$ reflects the coordinates indicated and $\Omega$ is the worldsheet 
parity. 
 
The splitting of the D6-branes into half-D6-branes discussed in  
\cite{karch} becomes obvious after T-duality. Indeed, it follows easily from  
the above formulas that the location of the upper half 6-branes,
$X_4=X_5=0, X_7>0$ in the type IIA configuration maps to the locus $z_2=0$ in
IIB. Similarly, the lower halfs of the 6-branes, $X_4=X_5=0, X_7<0$,
map to $z_1=0$. Thus the upper halfs of D6-branes map to {\it whole} 
D7-branes located at $z_2=0$, while the lower halfs map to {\it whole}
D7-branes at $z_1=0$. 
 
To specify the theory on the 7-branes completely we need to make a  
consistent choice for the action of the orientifolds on the Chan-Paton factors 
of the 7-7, 7-7$'$, and 7$'$-7$'$ strings. There are at least two such choices. 
One gives rise to an $SO(8)\times SO(8)$ gauge symmetry \cite{bz}, and
classically the 
other yields a $U(4)\times U(4)$ gauge group on the 7-branes \cite{sen1,sen2},
which is broken to $SU(4)\times SU(4)$ by one-loop effects~\cite{berkoozetal}. 
The second case is related to the Gimon-Polchinski \cite{gp} orientifold via
T-duality. We will be mainly interested in the second orientifold, which we
will refer to as the 
Sen model. Both of these orientifolds were constructed as compact models 
with a total of four orientifolds and sixteen physical 7-branes of each kind. 
The 7-brane gauge groups listed here are the parts of the total 7-brane 
group that are visible to a 3-brane probe near one of the intersections. 
 
The theory on a 3-brane probe in the Sen model background was analyzed 
in \cite{ah1}. The gauge group, matter content, and the superpotential are in 
complete agreement with the theory we discussed in section \ref{conformal}. 
Thus we conclude that the IIA configuration with all 6-branes on top of the 
NS5-brane is T-dual to a local piece of the Sen model \cite{sen1,sen2}.  
As in the IIA description the flat directions of the field theory correspond
to motions of the 3-branes in the 7-brane background. Moving the 3-branes
off the intersection point along either of the O7-planes corresponds to giving
en expectation value to one of the bifundamentals ${\cal Q},\tilde{\cal Q}$,
and moving the 3-branes off both orientifolds gives an expectation value to
both ${\cal Q}$ and $\tilde{\cal Q}$. Separating the
3-branes in the direction which the 7- and 7$'$-branes share corresponds to
giving expectation values to the antisymmetric tensors $A_1,A_2$.

It is instructive to study the deformations 
of the Sen model and compare these to the deformations of the 
corresponding IIA construction. The IIA construction has the advantage that 
all deformations of the background correspond to moving the 6-branes or the NS5-branes. In 
the IIB picture only some of the deformations are geometric, others correspond 
to Wilson lines. Once the map between IIA and IIB deformations is established, 
we can also find the IIB description of the second (non-conformal) IIA configuration  
discussed in section \ref{iic}.  
 
Sen \cite{sen1,sen2} has studied the deformations of the compact model in 
great detail. In the compact case the field theory on the 7-branes turns out  
to be a $(1,0)$ theory in six dimensions. Since our IIB configuration is non-compact,  
we cannot simply use Sen's results. 
In fact, in our case the theory on the 7-branes is not even six-dimensional, instead  
it is an eight-dimensional theory with six-dimensional 
impurities. Such theories have been discussed previously~\cite{kap2,kapdn}. 
 
Before we launch into an analysis of the impurity theory we need to discuss 
the matter content of the 7-brane theory. A single O7$^-$-plane with four 
coincident 7-branes gives rise to an ${\cal N}=1$ $SO(8)$ theory in eight dimensions.  
The bosonic degrees of freedom in the eight-dimensional vector multiplet consist of a 
vector field and a complex scalar, both in the adjoint of the gauge group. 
The second O7$^-$-plane in our configuration breaks half of the supersymmetries 
and imposes projections on fields in the vector multiplet. With the  
projection matrices for the Sen model \cite{gp,sen1}, the surviving constant modes of 
the fields are a vector and a complex scalar in the ${\bf 6}+ 
\bar{\bf 6}$. These fields account for the 7-7 strings and there are similar fields on  
the 7$'$-branes from 7$'$-7$'$ strings. The 7-7$'$ strings are localized at the intersection  
of 7- and 7$'$-branes. They yield a single hypermultiplet of the six-dimensional $(1,0)$  
theory on the intersection, which transforms as a $({\bf 4},{\bf 4})$ under the 
(classical) $U(4)_7\times U(4)_{7'}$ gauge group.
 
\subsection{The seven-brane impurity theory} 
 
In this section we analyze the supersymmetric vacua of the impurity theory on
the 7-branes and compare them with the vacua of the T-dual IIA configuration. 
We expect the vacuum field configurations to be translationally invariant in
the six directions common to the 7- and 7$'$-branes. Focusing now on the
7-branes,
we see that we can capture the physics by studying the dependence of the 
7-brane fields on the remaining two directions transverse to the 7$'$-branes.  
The 7$'$-branes and the O7$'$-plane intersect this two-dimensional plane 
in a point.  
To set up the impurity theory we use a complex affine coordinate $z$ on the plane 
and define $A_{\bar z} = \frac{1}{2}(A_1+iA_2)$, where $A_i$ are the two 
components of the $SO(8)$ gauge field living on the 7-branes. 
The 7-brane theory also contains a complex scalar, $\Phi$, in the adjoint of
$SO(8)$ that describes the transverse fluctuations of the 7-branes.
The bifundamental $({\cal M},\tilde{\cal M})$ from the 7-7$'$ strings is localized 
at the point $z=0$. A very similar theory (without orientifold projections)
was described in \cite{kap2}. The moduli space of the  
impurity theory is given by the solution of the equations 
\begin{eqnarray}\label{imp} 
&F_{z\bar{z}}-[\Phi,\Phi^\dagger] = 
\delta(z)\left({\cal MM}^\dagger  
-\tilde{\cal M}^\dagger\tilde{\cal M}\right)&\nl 
&\bar{D}\Phi=-\delta(z) {\cal M}\tilde{\cal M},& 
\end{eqnarray} 
where $F_{z\bar{z}}= \partial A_{\bar z} - {\bar \partial}A_z-[A_z,A_{\bar z}]$
and ${\bar D}={\bar\partial}-A_{\bar z}$.
These equations are known as Hitchin equations with sources. They are  
analogous to the $D$ and $F$ flatness conditions in ordinary supersymmetric field 
theories. A similar set of equations describes the impurity theory on the 
7$'$-branes.  
 
To make contact with the notation in \cite{sen1,sen2} we write all 7-brane
fields as antisymmetric $8\times 8$ matrices with certain constraints on the entries.  
This reflects the origin of the fields in the impurity theory. Without the 
O7$'$-plane, both $A_{\bar z}$ and $\Phi$ would transform in the adjoint of 
$SO(8)$. Orientifolding with O7$'$ puts additional constraints on these fields 
 
\begin{eqnarray}\label{constr} 
\Phi(z) &=& P\Phi^T(-z)P^{-1} \nl 
A(z) &=& PA^T(-z)P^{-1}, 
\end{eqnarray} 
where  
\begin{equation} 
P = \left( \begin{array}{cc} P_4 & 0 \\ 0 & -P_4 \end{array} \right), \qquad 
P_4 = \left( \begin{array}{cc} 0 & {\bf 1}_{2\times 2} \\ 
-{\bf 1}_{2\times 2} & 0 \end{array} \right). 
\end{equation} 
Orientifolding also breaks the gauge group from $SO(8)$ down to the group of all
continuous $SO(8)$-valued functions satisfying $g(z)=P g(-z) P^{-1}$. In particular,
at $z=0$ the gauge group reduces to $U(4)$.
The orientifold projections allow the bifundamentals to be arbitrary complex 
$8\times 8$ matrices that commute with $P$ \cite{gp}.
The impurity equations are consistent if the products of the bifundamentals
on the r.h.s.~of \eq{imp} are antisymmetrized in the gauge indices. 

We need to find all, possibly $z$-dependent, field configurations that satisfy 
the impurity equations, \eq{imp}, modulo gauge transformations. To this end we make 
the following ansatz 
\begin{equation} 
A_{\bar z} = \frac{T}{z}, \qquad \Phi(z) = \Phi_0 + \frac{\Phi_s}{z}. 
\end{equation} 
Here $T, \Phi_0$ and $\Phi_s$ are constant antisymmetric $8\times 8$ matrices. 
Imposing the constraints, \eq{constr}, determines that $\Phi_0$ transforms in
the ${\bf 6}+\bar{\bf 6}$ of $U(4)$ while $T$ and  $\Phi_s$ transform as 
adjoints. The background gauge field, $A_{\bar z}$, can be interpreted as a 
flat connection that gives rise to a Wilson line around the intersection point at 
$z=0$. The constant part of the scalar field, $\Phi_0$, corresponds to the asymptotic
(i.e. $z\to\infty$) positions of the 7-branes in the directions transverse to the 
O7-plane, while the singular part, $\Phi_s$, parametrizes a deformation of the shape of 
the 7-branes. 
 
The moduli space of the impurity equations, \eq{imp}, has several branches 
with rather different physics. The simplest situation arises if all 
bifundamental expectation values and all singular parts of $A_{\bar z}$ and $\Phi$ 
vanish. In that case \eq{imp} reduces to the condition 
\begin{equation} 
[\Phi_0,\Phi_0^\dagger] = 0, 
\end{equation} 
which is solved by 
\begin{equation}\label{six} 
\Phi_0 = \left( \begin{array}{cc} 0 & \phi \\ -\phi & 0 \end{array}\right), 
\qquad 
\phi = \diag(\phi_1,\phi_2,\phi_1,\phi_2). 
\end{equation} 
As in Ref.~\cite{sen1}, the two complex parameters, $\phi_{1,2}$, parametrize
the transverse position of two pairs of 7-branes. We discuss the corresponding
IIA deformation in the next section. For the remainder of this section we
set $\Phi_0=0$.
 
The impurity equations, \eq{imp}, become inhomogeneous once we turn on an
expectation value for the bifundamental fields. Since 
${\bar\partial}(1/z) \sim \delta(z)$, and the r.h.s.~of \eq{imp} is
proportional to $\delta(z)$, the singular fields above have the  
right form to satisfy the impurity equations with nonzero bifundamental 
expectation values.

The most generic expectation value of the bifundamentals for which the 
impurity equations have solutions reads 
\begin{eqnarray} 
&{\cal M} = \left(\begin{array}{cc} M_1 & 0 \\ 0 &  M_2 \end{array} \right),&\\ 
& M_1 = \left( \begin{array}{cccc} 
	m_1 & 0 & -im_1 & 0 \\ 
	0 & m_2 & 0 & -im_2 \\ 
	im_1 & 0 & m_1 & 0 \\ 
	0 & im_2 & 0 & m_2 \end{array} \right), \qquad 
M_2 = \left( \begin{array}{cccc} 
        m_3 & 0 & im_3 & 0 \\ 
        0 & m_4 & 0 & im_4 \\ 
        -im_3 & 0 & m_3 & 0 \\ 
        0 & -im_4 & 0 & m_4 \end{array} \right),&\nonumber 
\end{eqnarray} 
and an expectation value of the same form, but with $m_i$ replaced by $\tilde{m}_i$,
for $\tilde{\cal M}$.
The impurity equations determine the expectation values of the other fields
in terms of ${\cal M}$ and $\tilde{\cal M}$. The residue of $\Phi$ 
is given by
\begin{equation}\label{x45}
\Phi_s = \diag(\Phi_1,\Phi_2),
\end{equation}
where
\begin{equation}
\Phi_1 = \left( \begin{array}{cccc} 
        0 & 0 & -\phi_1 & 0 \\ 
        0 & 0 & 0 & -\phi_2 \\ 
        \phi_1 & 0 & 0 & 0 \\ 
        0 & \phi_2 & 0 & 0 \end{array} \right), \qquad 
\Phi_2 = \Phi_1(\phi_1\to-\phi_3,\phi_2\to -\phi_4),
\end{equation} 
with $\phi_i \sim m_i\tilde{m}_i$. The matrix $T$ in the gauge connection has
the same structure as $\Phi_s$, except that $\phi_i$ is replaced by 
$t_i \sim |m_i|^2 - |\tilde{m}_i|^2$.

Before discussing this general solution, we will focus on two special cases.
If we set $m_i = \tilde{m}_i$, the r.h.s.~of the first impurity equation 
vanishes and only the residue of $\Phi$ is turned on.
This expectation value of the bifundamentals
breaks the $U(4)\times U(4)$ gauge group
to a diagonal 
subgroup. If all $m_i$ are equal this subgroup is $U(4)_D$, and for generic 
values of $m_i$ we find $U(1)^4$. Since the 
7-brane group is broken to a diagonal subgroup, the impurity theory, 
\eq{imp}, on the 7-branes and the corresponding impurity theory on the 7$'$-branes
contain the same information. Therefore it is sufficient to consider only the
7-brane theory. The field $\Phi(z)$ describes the 
shape of the 7-branes. For large $z$ the 7-branes asymptote to the O7-plane
as in the unperturbed case, while they approach the O7$'$-plane for small $z$.
Thus we conclude that turning on this bifundamental expectation value deforms
pairs of intersecting 7- and 7$'$-branes into a single smooth 7-brane that
interpolates between the 7- and 7$'$-branes. 
This result agrees with the F-theory analysis in \cite{sen2}, where this 
behavior was interpreted as fusing the 7- and 7$'$-branes together.
 
There are also solutions of the impurity equations with non-zero gauge  
connection and $\Phi_s=0$. We find one such solution if we set 
$m_1 = m_2=\tilde{m}_3=\tilde{m}_4$,
and all other components of the bifundamentals vanish. For this choice the  
r.h.s.~of the second equation in \eq{imp} vanishes, which implies $\Phi_s=0$, 
and $t_1=t_2 \sim |m_1|^2$, $t_3=t_4 \sim -|m_1|^2$.
This bifundamental expectation value breaks the $U(4)_7\times U(4)_{7'}$ 
7-brane gauge group to a diagonally embedded $U(2)\times U(2)$.  
Note that this deformation is purely non-geometric. Since $\Phi(z)=0$, the
7-branes have the same shape as in the case without any bifundamental
expectation values.

It is now a simple matter to identify these two singular solutions with the 
corresponding deformations in the IIA construction. The first solution with
$T=0$, $\Phi_s \neq 0$  
corresponds to moving the 6-branes off the NS5-brane in the $X_{4,5}$ 
direction. If none of the 6-branes coincide, the $U(4)\times U(4)$ gauge 
symmetry on the 6-branes is broken to $U(1)^4$. This is in complete agreement 
with the impurity analysis. Note 
that a deformation that corresponds to fusing 7 and 7$'$-branes together in  
the IIB description maps into a simple brane motion in the IIA construction,
which involves reconnecting the upper and lower halfs of the 6-branes.

The second singular solution with $T\neq 0$, $\Phi_s = 0$
also corresponds to a simple brane  
motion in the IIA description. We identify turning on $m_1$ 
with the motion of two pairs of 6-branes in the $X_6$ direction.  
The classical gauge group on the 6-branes is $U(2)\times U(2)$ as expected
from the
IIB analysis.  This brane motion also requires that we reconnect the upper and
lower halfs of the 6-branes, so that the resulting 6-brane group is a diagonal
subgroup of the original $U(4)\times U(4)$ gauge symmetry. This is in
perfect agreement with the analysis of the 7-brane impurity theory. 

It is straightforward to discuss more general choices for the bifundamental
expectation values. The bifundamental expectation values are parametrized
by eight complex numbers, $m_i$ and $\tilde{m}_i$, which determine the
matrices $T$ and $\Phi_s$ completely. The four parameters in $T$ map into
the $X_6$ position of the 6-branes in the IIA description and the entries
in $\Phi_s$ correspond to the $X_{4,5}$ positions. Thus we find complete
agreement between the brane motions in the IIA description and the moduli
corresponding to singular fields in the impurity theory.
 
\subsection{A supersymmetric IIA configuration with curving six-branes} 
 
\begin{figure} 
\centerline{\epsfxsize = 5truecm \epsfbox{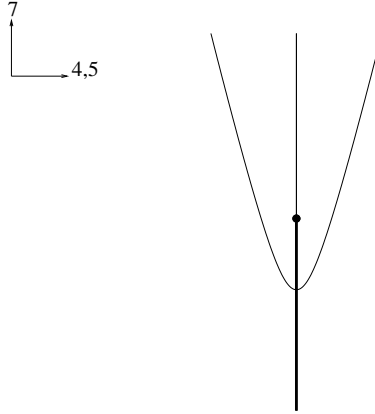} } 
\tighten{ 
\caption{ Type IIA configuration for non-zero expectation value of the {\bf 6}. 
The dot represents the NS5-brane, the thick line corresponds to four half  
6-branes, the thin line corresponds to two half 6-branes and the curving line 
is another 6-brane.  } 
\label{fig2} 
} 
\end{figure}

The deformations we discussed so far are rather complicated in the IIB 
picture and correspond to simple brane motions in the IIA description.  
In fact, all simple brane motions in the IIA description are accounted for.
However, there is 
a very simple brane motion in IIB, namely the constant solution of the impurity equations 
given in \eq{six}, that should have a counterpart in the IIA description.  
Since this deformation corresponds to moving pairs of 7-branes off the 
orientifold, we can find an explicit equation describing the position of these 
branes. In terms of the coordinates in \eq{z12} this equation reads 
$z_2 = const$.  Starting from this expression 
we can reverse the coordinate transformations that took us from the Taub-NUT 
space to the flat coordinates on ${\bf C}^2/{\bf Z}_2$. This provides an 
expression for the world volume of the 7-brane in the Taub-NUT coordinates. 
Since the 7-branes wrap the fiber of the Taub-NUT and the fiber T-dualizes to 
the compact $X_6$ direction, it is straightforward to find the equation for the 
worldvolume of the corresponding 6-brane. The result is $X_4^2-c X_7-c^2/4=0$, i.e., a 
parabola in the $X_4-X_7$ plane. Fig.~\ref{fig2} shows the IIA  
configuration which is T-dual to the following IIB situation: all 7$'$-branes are 
coincident with the O7$'$-plane, and one pair of 7-branes is displaced from   
the O7-plane. 
 
From this picture one can see that turning on the constant complex scalar on the 7-brane  
corresponds to fusing two upper halfs of the 6-branes together and moving them off the  
NS5-brane as shown in the figure. On the IIB side it is obvious that
this deformation preserves all supersymmetries. This is somewhat harder to see on the IIA side.
Presumably the $H$-field produced by the NS5-brane stabilizes the curved worldvolume of the
D6-brane.

The effect of this deformation on the probe theory is what we expect from 
the IIB picture. There we move two 7-branes away from the 3-branes sitting at
the intersection point of the orientifold planes. This gives a mass to half of
the fundamentals from 7-3 
strings. In the IIA picture the deformation accomplishes the same. In the 
IIB picture moving the 3-branes along the O7$'$-plane and transverse to the
O7-plane corresponds to 
giving the bifundamental field ${\cal Q}$ in the probe theory an expectation value 
\cite{ah1}. Thus it is possible to move the 3-branes away from the intersection of the 
orientifolds towards the intersection of the pair of 7-branes with the O7$'$-plane by giving 
an expectation value to one of the bifundamentals. This is also reflected in the  
IIA description. We can move the 4-branes in the negative $X_7$ direction by giving an  
expectation value to one of the bifundamentals (see section \ref{iib}). This moves the  
4-branes off the NS5-brane and towards the intersection of the lower half-6-branes with  
the curving 6-brane. 
 
In the IIB description moving a pair of 7-branes away from the O7-plane breaks the  
7-brane gauge group from $SU(4)$ down to $SU(2)\times SU(2)$~\cite{sen1,sen2}.  
Moving all four 7-branes together breaks $SU(4)$ down to $Sp(4)$. This implies that 
the unbroken gauge group on a single curving 6-brane should be $SU(2)$, 
while for two coincident curving branes it should be enhanced to $Sp(4)$. 
It is not at all clear how to see this from the IIA description.  
 
\subsection{Comparison with F-theory} 
 
Sen argued \cite{sen1,sen2} that the T-dual version of the GP model  
\cite{gp} is related to an F-theory compactification with certain fluxes through 
collapsed 2-cycles. The naive candidate for such an F-theory compactification 
would be a pair of intersecting $D_4$ singularities. However, this cannot be 
directly related to the GP orientifold, since it would give rise to an  
$SO(8)\times SO(8)$ gauge symmetry and contain tensionless strings, while the GP model  
has $SU(4)\times SU(4)$ symmetry and no tensionless strings. The difference is due to  
NS (and possibly RR) 2-form fluxes through the collapsed 2-cycle at the intersection of the 
two $D_4$ singularities. These fluxes give a mass to 3-branes wrapping this 
cycle, thereby preventing the appearance of tensionless strings. These fluxes 
are not quantized \cite{sen1,sen2}, so we should be able to identify moduli 
in our IIA description that correspond to turning them off. The NS flux is 
conventionally identified with the position of the NS5-branes on the $X_6$ 
circle and the RR flux parametrizes the location of the NS5-branes on the  
M-theory circle. From the IIB point of view, they are both part of a massless 
hypermultiplet living at the intersection of the $D_4$ singularities. In order to turn 
off the NS flux, we move the NS5-brane and its image as well as all D6-branes to 
coincide with one of the O6-planes. 
This configuration has an $SO(8)\times SO(8)$ gauge symmetry from 
the eight upper and eight lower halfs of the 6-branes, as well as tensionless strings from  
the NS5-brane coincident with its image \cite{hz}. 
In addition to the hypermultiplet that corresponds to moving the NS5-brane 
off the orientifold in the $X_4,X_5,X_6,X_{10}$ there is now a tensor multiplet whose 
scalar expectation value corresponds to separating the two NS5-branes in the $X_7$ direction.  
All this agrees with the expectations from F-theory.  
 
\section{The large $N$ limit} 
\label{sugra} 
 
When the number of D3-branes, $N$, is large there is a dual description of
${\cal N}=1$ superconformal theory on the D3-branes in terms of a supergravity
on $AdS_5\times X$,
where $X$ is an Einstein manifold (or orbifold)~\cite{m1}. 
This dual description is valid when the t'Hooft gauge coupling, $g_{YM}^2 N$,
is large. In this section we will show how the AdS/CFT correspondence works  
for the conformal gauge theory with $SU(4)\times SU(4)$ flavor symmetry discussed in 
section \ref{iib}, and provide evidence that this theory is finite. We will also provide 
a partial analysis of the non-conformal theory of section \ref{iic} in the large $N$ limit 
and argue that supergravity suggests a definite R-charge assignment for all the fields 
in the infrared. 
 
\subsection{The conformal case} 
 
In the conformal case, $X$ is an orientifold of ${\bf S}^5$. As explained in
the previous section, the IIA configuration with $SU(4)\times SU(4)$ gauge
symmetry on the 6-branes is T-dual to a local piece of the Sen model. At the
$SU(4)\times SU(4)$ point, the Sen model is a perturbative type IIB  
orientifold with constant string coupling, $\tau$ \cite{sen1,sen2}. Thus the 
near-horizon geometry of the 3-branes is obtained by orientifolding $AdS_5\times {\bf S}^5$.  
Similar theories were analyzed in \cite{f1,f2,kap}. 

Let us denote the orientifolded five-sphere by ${\tilde{\bf S}}^5$.  
The metric on ${\tilde{\bf S}}^5$ is the angular part of 
\begin{equation} 
ds^2 = |dz_1|^2 + |dz_2|^2 + |dw|^2, 
\end{equation} 
where $w=X_8+iX_9$ and the variables $z_1,z_2$ are subject to the
identifications $z_i \to -z_i$. A $U(1)^3$  subgroup of the $SO(6)$ isometry
group of ${\bf S}^5$ commutes with these identifications. It is convenient
to take the generators that rotate $z_1$, $z_2$,and $w$ separately as a basis
in the Lie algebra of $U(1)^3$. Explicitly, the metric on ${\tilde{\bf S}}^5$
can be written as  
\begin{equation} 
ds^2_{{\tilde{\bf S}}^5} = d\theta_1^2+\sin^2(\theta_1)d\phi_1^2 + \cos^2(\theta_1) 
\left( d\theta_2^2+\sin^2(\theta_2)d\phi_2^2+\cos^2(\theta_2)d\phi_3^3\right), 
\end{equation} 
where $\phi_{1,2} \in [0,\pi]$, $\phi_3 \in [0,2\pi]$, and 
$\theta_{1,2} \in [0,\pi]$. The three angles $\phi_i$ parametrize rotations 
in the $z_{1,2}$ and $w$ planes respectively. The periodicity of $\phi_{1,2}$ 
reflects the identifications on $z_{1,2}$. Since this periodicity of $\phi_{1,2}$  
is the only thing which distinguishes ${\tilde{\bf S}}^5$ from ${\bf S}^5$, the eigenvalues 
of the scalar Laplacian on the former can be obtained from those on the latter. 
The eigenvalue of the scalar Laplacian on ${\bf S}^5$ is $k(k+4)$, where $k=0,1,\ldots$.  
In terms of the angular momenta, $m_i$, associated with the angles $\phi_i$,
we have 
$k = |m_1|+|m_2|+|m_3|+2l_1+2l_2$, where $l_i$ are non-negative integers.
The orientifold projection on the bulk supergravity states amounts to  
keeping modes with even $m_1$ and $m_2$.  

In the ${\cal N}=4$ case, the supergravity states with lowest mass squared
come from the KK reduction of $h^a_a$, the dilaton mode of the
${\bf S}^5$. The AdS masses of these states are given by $m^2 = k(k-4)$ \cite{romans},
where $k$ is given above. According to \cite{w1}, the AdS mass of a KK state is
related to the dimension of the corresponding boundary operator by
$\Delta(\Delta-4)=m^2$, which implies $\Delta=k$ for this tower of KK modes. 
The decomposition of the other supergravity fields yield towers of KK states
for which $\Delta = k+n$, where $n$ is a positive integer \cite{w1}.
We will see below that only for $n=0$ the KK states couple to chiral primary operators. 
Therefore we will restrict our analysis to the KK modes from the 
decomposition of $h^a_a$.

The simplest way to identify chiral primaries is to find all states for which 
$\Delta=\frac{3}{2}R$, where $R$ is the R-charge which is part of the superconformal algebra. 
The R-current is a certain linear combination of the three $U(1)$ currents. 
To find this linear combination we first need to determine which supercharges 
survive the orientifold projection. The orientifold group, ${\bf Z}_2\times
{\bf Z}_2$, is generated by $\gamma_1=R_{z_2}\Omega (-1)^{F_L}$ and
$\gamma_2=R_{z_1z_2}$.  
Orientifolding by the first generator breaks $SO(6)$ down to $SU(2)_L\times SU(2)_R 
\times U(1)_N$ where $U(1)_N$ acts on $z_2$ while $SU(2)_L\times SU(2)_R$ acts 
on $z_1,w$. The surviving supercharges $(Q_+,Q_-)$ transform as $({\bf 1},{\bf 2})_1$
with respect 
to this group. Orientifolding by $\gamma_2$ breaks $SU(2)_L\times SU(2)_R$ down to 
$U(1)_L\times U(1)_R$. We will denote the sum of the $U(1)_L$ and $U(1)_R$
charges 
by $U(1)_2$, the difference by $U(1)_3$, and refer to $U(1)_N$ as $U(1)_1$. 
The charges of $z_2, z_1,$ and $w$ under these three $U(1)$'s are given by 
$(2,0,0),(0,2,0),$ and $(0,0,2)$, respectively. The supercharge $Q_+$ which survives the 
second orientifolding has $U(1)$ charges $(1,1,-1)$. It follows that the R-charge which 
is in the same superconformal multiplet as the stress-energy tensor is  
$\frac{1}{3}(2m_1+2m_2-2m_3)$. Here $2m_1$ is the $U(1)_1$ charge, $2m_2$ is the  
$U(1)_2$ charge, and $2m_3$ is the $U(1)_3$ charge. 
The normalization is chosen so that $Q_+$ has R-charge $1$. 
It follows that any KK mode with 
$l_1=l_2=0$, $m_1,m_2 \ge 0$ and $m_3 \le 0$ should couple to a chiral 
primary operator in the boundary field theory.  
 
We discussed the identification of geometric motions of 3-branes with flat
directions in the 3-brane field theory in the previous section (see also
\cite{ah1}). This allows us to
determine the $U(1)$ charges of the fields $A_1,A_2,{\cal Q},\tilde{\cal Q}$.
The field theory superpotential then fixes the R-charges of the fundamentals 
$q,{\tilde q},p,{\tilde p}$. The results are summarized in the table below. 
 
\begin{equation}\label{table} 
\begin{array}{c|ccc} 
& U(1)_1 & U(1)_2 & U(1)_3 \\ \hline 
A_{1,2} & 0 & 0 & -2 \\ 
{\cal Q} & 2 & 0 & 0 \\ 
\tilde{\cal Q} & 0 & 2 & 0 \\ 
q,p & 0 & 1 & -1 \\ 
\tilde{q},\tilde{p} & 1 & 0 & -1 
\end{array} 
\end{equation} 
 
With these charge assignments in hand it is now a simple matter to match 
the bulk KK modes and the chiral primary operators in the field theory. Let us 
give some examples. 
The supergravity spectrum contains a singleton chiral primary with  
$U(1)_3$ charge $-2$ and $\Delta =k=1$. 
This state corresponds to a chiral primary $\Tr (A_1 J_1) + \Tr (A_2 J_2)$ in the field  
theory.\footnote{The antisymmetric representation of $Sp(N)$ is reducible and  
contains a singlet.} Since $\Delta=1$, this is a free field. For $\Delta = 2$ 
there are three chiral primary states with geometric $U(1)$ charges $(4,0,0)$, $(0,4,0)$, 
and $(0,0,-4)$. We identify them with $\Tr{\cal Q}^T J_1 {\cal Q} J_2$, 
$\Tr\tilde{\cal Q}^T J_2 \tilde{\cal Q} J_1$, and $\Tr (A_1 J_1)^2 + \Tr (A_2 J_2)^2$.  
The chiral primary operators with $\Delta = 3$ are
$\Tr[{\cal Q} A_2 {\cal Q}^T J_1+{\cal Q}^T J_1 A_1 J_1 {\cal Q} J_2]$, 
$\Tr[\tilde{\cal Q} A_1 \tilde{\cal Q}^T J_2+
\tilde{\cal Q}^T J_2 A_2 J_2 \tilde{\cal Q} J_1]$, and  
$\Tr (A_1 J_1)^3 + \Tr (A_2 J_2)^3$. They correspond to the KK states with 
charges $(4,0,-2)$, $(0,4,-2)$, and $(0,0,-6)$ respectively. 
 
The field theory also contains operators that carry charges under the  
7-brane gauge groups. It was pointed out in \cite{f2} that these operators 
couple to the AdS modes coming from the KK reduction of the 7-brane fields. 
Our configuration includes an O7-plane with four coincident D7-branes  
wrapping an ${\bf S}^3$ defined by $|z_1|^2+|w|^2=const$,
and similarly an O7$'$-plane 
with four D7$'$-branes wrapped on $|z_2|^2+|w|^2=const$. 
The two 3-spheres intersect over a circle. 
We can focus on the KK modes from the first ${\bf S}^3$. These modes couple  
to operators that are charged under the $SU(4)_7$ subgroup of the
$SU(4)_7\times SU(4)_{7'}$ 
global symmetry group of the probe theory. The modes living on the other
${\bf S}^3$ couple to similar operators in the field theory that transform
under $SU(4)_{7'}$. 
 
The KK reduction of the theory on an O7-plane with four coincident 7-branes 
was discussed in \cite{f2}. In that case there were twice as many supersymmetries as in 
ours. The simplest way to compute the KK spectrum in our case is to 
use the results of \cite{f2} and impose the additional projection from the 
O7$'$-plane.   
 
Ref.~\cite{f2} contains a detailed discussion of the 7-brane states and 
their multiplet structure. The lowest component of the multiplet is a  
real field in the $({\bf k}, {\bf k+2})_0$ representation of  
$SU(2)_L\times SU(2)_R\times U(1)_N$, where $k=1,2,\ldots$. 
This mode comes from KK reduction of the components of the 7-brane gauge 
field along the ${\bf S}^3$,  
\begin{equation} 
A_a = \sum_k a_k Y_a^k, 
\end{equation} 
where $Y_a^k$ is the $k$-th vector spherical harmonic on ${\bf S}^3$. These modes 
couple to operators of dimension $\Delta = k+1$ in the boundary field theory. 
For simplicity we will only consider operators with $\Delta = 2,3$.  
The state with $\Delta=2$ transforms
in the $({\bf 1},{\bf 3})_0$ and decomposes into 
modes with $U(1)^3$ quantum numbers $(0,0,0)$ and $(0,\pm 2, \mp 2)$.  
The $(0,0,0)$ mode has no $U(1)_R$ charge and does not correspond to a chiral primary.  
The states with $U(1)^3$ charges $(0,2,-2)$ and $(0,-2,2)$ are complex
conjugates of each other, so it is sufficient to consider only one of them,
e.g., the first. It has R-charge $4/3$ and is, therefore, a chiral primary. 
This state starts out in the adjoint of the $SO(8)_7$ gauge group on the
7-brane. Since it has $m_2=1$, it is odd under the additional orientifold
projection $\gamma_2$. This projection breaks $SO(8)_7$ down to $SU(4)_7$. As
explained in~\cite{sen1,sen2}, states in the adjoint of $SO(8)_7$ which are
odd under $\gamma_1$ yield ${\bf 6}+\bar{\bf 6}$ of $SU(4)_7$, while even states
give adjoints of $SU(4)_7$. It follows that the 
$(0,2,-2)$ state yields one complex state in ${\bf 6}$ and one complex state  
in $\bar{\bf 6}$. These KK states correspond to operators $qJ_1q$ and $pJ_2p$, which  
transform in the $\bf 6$ and $\bar{\bf 6}$ of the 7-brane group respectively. 
 
The $\Delta=3$ mode is in the $({\bf 2},{\bf 4})_0$ representation and decomposes  
into even modes with $U(1)^3$ charges $(0,0,-2)$ and $(0,4,-2)$ and their complex  
conjugates, as well as odd modes with $U(1)^3$ charges $(0,2,0)$ and $(0,2,-4)$ and  
their complex conjugates. 
The even $(0,0,-2)$ mode and the odd $(0,2,0)$ mode do not couple to chiral primary operators, 
because the R-charge does not match the dimension. The even $(0,4,-2)$ 
mode couples to a chiral primary operator in the adjoint of $SU(4)$ which we 
identify as $pJ_2\tilde{\cal Q}J_1q$. The odd $(0,2,-4)$ mode couples to a 
chiral primary in the ${\bf 6}+\bar{\bf 6}$. The corresponding operators are
given by $qA_1q$ and $pJ_2A_2J_2p$.  
 
Other scalars on AdS come from the decomposition of the complex scalar 
field on the 7-branes. These KK modes are in the $({\bf k},{\bf k})_2$ 
representation of $SU(2)_L\times SU(2)_R\times U(1)_N$ and couple to operators of  
dimension $k+2$~\cite{f2}. It is straightforward to decompose and project these modes as we 
did for the KK modes of the vector field. The $\Delta=3$ case is especially 
simple, since this mode carries only $U(1)_1$ charge. Since the R-charge 
and the dimension do not satisfy $\Delta=\frac{3}{2} R$, this KK mode does not  
couple to a chiral primary operator. The same is true for the higher KK modes of  
the complex scalar field. 
 
Finally, there are also states living on the intersection of the 7-branes and 7$'$-branes 
which is an ${\bf S}^1$ embedded in ${\bf S}^5$. The KK reduction of these states is  
straightforward, and we will not discuss it. 
 
In the above analysis we have focused on chiral primaries. It is also interesting 
to ask whether non-chiral states match between field theory and supergravity. 
Some of the non-chiral scalars we have seen, namely the ones coming from the 
reduction of complex scalars living on the 7-branes, are descendants of the 
chiral primaries and therefore match automatically. On the other hand, the 
non-chiral scalars which come from the KK reduction of the gauge field on the 7-branes 
are primary. One may ask whether the superconformal multiplet they live in is 
long or short. 

To answer this question we need to recall some
facts about unitary representations of the ${\cal N}=1$ superconformal algebra~\cite{conf}.
For our purposes it is sufficient to consider multiplets whose primary states have zero spin.
Let the $R$ and $\Delta$ be the R-charge and the dimension of the primary. Unitarity
puts restrictions on which values of $R$ and $\Delta$ may occur; the allowed possibilities
are

(i) $\Delta=R=0$ (the trivial representation),

(ii) $\Delta=\frac{3}{2}|R|$ (chiral and anti-chiral representations),

(iii) $\Delta\geq\frac{3}{2}|R|+2$.

Representations of type (iii) with $\Delta>\frac{3}{2}|R|+2$ contain no null states
and therefore are termed long multiplets. Chiral and anti-chiral representations contain 
null states at level one, i.e., their primaries are annihilated by half of the supercharges.
These representations are called short.
Representations of type (iii) which saturate the inequality are also short; the null states
occur at level two. A well-known example of a short multiplet is a linear multiplet which 
contains a conserved current. It corresponds to the case $R=0,\Delta=2$. 

One can check that all non-chiral primaries coming from the reduction of the gauge field
on the 7-branes satisfy $\Delta=\frac{3}{2}|R|+2$ and therefore are in short multiplets
of type (iii).
In particular, the $(0,0,0)$ mode with $\Delta=2$ we have found above 
is in fact the lowest component of a linear multiplet. It couples to a field theory operator  
$q^\dagger q -  pp^\dagger$ in the adjoint of $SU(4)_7$. The corresponding current  
is simply the $SU(4)_7$ flavor current. 
The matching of non-chiral primaries with $\Delta=3$ is a bit more involved.  
The $(0,2,0)$ mode transforms in ${\bf 6}+\bar{\bf 6}$ of $SU(4)_7$. Its
field theory  
counterparts are $h_1 p^\dagger\tilde{\cal Q}J_1 q + h_2 q J_1 A_1^\dagger J_1 q$ and 
$h_1 pJ_2\tilde{\cal Q} q^\dagger+h_2 p A_2^\dagger p$, where the flavor indices 
are antisymmetrized. The $U(1)^3$ charges of these operators match those of the
$(0,2,0)$ mode. To show that these operators live in short multiplets,
i.e., are annihilated by $\bar D^2$, one needs to use the classical equations
of motion.  The manipulations one has to go through are very similar to those
in~\cite{berkooz}, and are subject to the same caveats. The use of the classical equations of
motion is presumably justified in the weakly coupled regime where $g^2_{YM} N$ is small. The
supergravity analysis indicates that the operators in question belong to short
multiplets even for large $g^2_{YM} N$. An even more interesting 
situation arises when one tries to match the non-chiral primary with $U(1)^3$ charges 
$(0,0,-2)$ and $\Delta=3$. This mode lives in the adjoint of $SU(4)_7$. We
claim that it corresponds to the field theory operator 
$h_1 q A_1 J_1q^\dagger-h_1 p^\dagger A_2 J_2p-h_2 q \tilde{\cal Q}^\dagger p$. 
Evaluating the ${\bar D}^2$ descendant of this operator using the 
classical equations of motion, one finds that it does not vanish. Instead, the
descendant has the form $(q\tilde{q})(\tilde{p}p)$, i.e., it factorizes into a
product of two gauge-invariant operators and is therefore subleading at large
$N$. It follows that this field theory operator lives in a long 
multiplet for finite $N$, but is ``close'' to being in a short multiplet
in the sense that its dimension approaches the unitarity bound as $N\to\infty$. 
On the supergravity side this means that the $(0,0,-2)$ one-particle state is in a short 
multiplet only for $N=\infty$. For finite $N$ the multiplet absorbs another short
multiplet made of two-particle states and becomes long.

This concludes our analysis of the AdS/CFT correspondence for the Sen model.  
There is complete agreement between the spectrum of primary operators 
in the field theory and the scalar Kaluza-Klein states on AdS as required 
by the AdS/CFT correspondence~\cite{m1,gubs,w1}. The charge assignments in
Table~\ref{table}
together with the formula $R=\frac{1}{3}(2m_1+2m_2-2m_3)$ imply that all  
chiral fields have canonical dimensions in the infrared. This is the most  
natural assumption for a theory with  
vanishing beta function, but as we pointed out in the introduction there is no 
field theory proof of this. The supergravity computation is only valid for 
large $N$ and large $g^2_{YM} N$. However, given that for $g^2_{YM}\ll 1$ and
$N$ of 
order $1$ the dimensions are also canonical, it appears likely that the theory is 
finite for all $N$. 
 
\subsection{The non-conformal case} 
 
Next we discuss the deformed ${\cal N}=1$ theory which flows to a line of conformal 
fixed points in the infrared (section \ref{iic}). We have already pointed out that 
although the Wilsonian gauge coupling in this theory depends on the scale, 
the low-energy effective gauge coupling does not vary over the moduli space. 
This implies that the corresponding IIB background should have constant 
$\tau$. Indeed, in section \ref{IIB} we showed 
that the 7-brane background for this configuration is very similar to the 
background for the conformal theory. As in the conformal case, the 7-branes 
do not bend and are coincident with the O7-planes. The RR charge of the 7-brane
is cancelled 
locally by the O7-planes, so we expect that the type IIB string coupling is constant. 
Similarly, the gravitational field of the 7-branes cancels against that of the 
orientifold planes. Thus it appears that the closed string sector is not
affected by this deformation. The only difference between the conformal and the 
non-conformal case is in the open string sector, namely in the gauge connection on the  
7-branes. In the conformal case it is trivial, while in the non-conformal 
case it is a flat connection which breaks the $SU(4)_7\times SU(4)_{7'}$ group to a  
diagonally embedded $SU(2)\times SU(2)$. To summarize, the deformation of the 7-brane 
background that leads to the non-conformal theory changes the properties of the 
theory on the 7-branes, but it appears not to change the closed string sector.  
 
To find a supergravity dual for this non-conformal theory, we need to repeat 
the analysis above with the new 7-brane background. Since the closed string sector 
is unchanged, the spectrum of the bulk modes should be the same as before. 
The matter content of the conformal and the non-conformal theory differ only 
in the number of flavors and their coupling to the bifundamentals. Therefore
both theories have the same spectrum of operators that do not transform under 
the 7-brane groups. Thus it appears that the dimensions of all chiral primaries 
uncharged with respect to the flavor group are the same as in the conformal case, 
i.e., canonical. If antisymmetric tensors and bifundamentals have zero anomalous 
dimensions then the vanishing of the beta-functions, \eq{beta2}, requires that  
the fundamentals have dimension $1/2$. This is actually the lowest dimension
for the fundamental allowed by unitarity. To show that this assignment of dimensions,  
or equivalently of R-charges, agrees with supergravity we would have to show that the KK  
reduction of the 7-brane theory with the singular flat connection switched on, reproduces 
the expected dimensions of the chiral primaries that involve the fundamentals.  
Unfortunately we do not know how to analyze the excitations of the impurity theory 
around nontrivial vacua, so we cannot check that our solution is consistent. Nevertheless,  
we get a definite prediction for the infrared dimensions of all fields. It would be  
interesting to confirm the answer by directly analyzing the perturbative expansion of the  
non-conformal theory at large $N$.

\acknowledgements 
It is a pleasure to thank O.~Aharony, E.~Gimon, J.~Maldacena, G.~Moore,
E.~Katz, and  E.~Witten 
for helpful discussions. M.G.~would like to thank the Institute for Advanced 
Study for hospitality while this work was in progress. 
The work of M.G~was supported in part by DOE grants \#DF-FC02-94ER40818 and
\#DE-FC-02-91ER40671, while
that of A.K. by DOE grant \#DE-FG02-90ER40542.

{\tighten

} 
\end{document}